\documentclass[aps,prd,preprint,tightenlines,groupedaddress,showpacs]{revtex4}
\usepackage{epsf,epsfig,graphics,graphicx}
\begin{document}

\title{Primordial black holes and associated gravitational waves in axion monodromy inflation}

\author{Shu-Lin Cheng$^1$}
\author{Wolung Lee$^1$}
\author{Kin-Wang Ng$^{2,3}$}

\affiliation{
$^1$Department of Physics, National Taiwan Normal University,
Taipei 11677, Taiwan\\
$^2$Institute of Physics, Academia Sinica, Taipei 11529, Taiwan\\
$^3$Institute of Astronomy and Astrophysics, Academia Sinica, Taipei 11529, Taiwan}

\date{\today}

\begin{abstract}
In the axion monodromy inflation, the inflation is driven by the axion with super-Planckian field values in a monomial potential with superimposed sinusoidal modulations. The coupling of the axion to massless gauge fields can induce copious particle production during inflation, resulting in large non-Gaussian curvature perturbation that leads to the formation of primordial black holes. In this paper, we explore the parameter space in the axion monodromy inflation model that favors the formation of primordial black holes with masses ranging from $10^8$ grams to $20$ solar masses. We also study the associated gravitational waves and their detection in pulsar timing arrays and interferometry experiments.
\end{abstract}

\pacs{98.80.Cq, 97.60.Lf, 04.30.-w}
\maketitle

\section{Introduction}

The inflation scenario is generally accepted for explaining the observed spatial flatness and homogeneity of the Universe.
A simple model of the scenario such as the slow-roll inflation driven by a flat inflaton potential predicts quasi de Sitter
vacuum fluctuations during the inflation which could give rise to Gaussian and
nearly scale-free metric perturbations containing both matter density fluctuations (scalar modes)
and gravitational waves (GWs or tensor modes)~\cite{olive}. 

The simplest model that provides the flat potential is perhaps the large-field inflation with a monomial potential. The drawback is that the inflaton needs 
super-Planckian field values to fulfill the slow-roll conditions. However, these large field values can be realized in the string theory and hence are utilized in the axion monodromy inflation~\cite{silverstein}. The inflaton potential in a single-field axion monodromy inflation has a monomial form with superimposed sinusoidal modulations whose size is model-dependent, given by~\cite{silverstein,mcallister,flauger}
\begin{equation}
V(\varphi)=V_0+\mu^{4-p} \varphi^p +\Lambda(\varphi)^4 \cos\left[\frac{\varphi}{f(\varphi)}+\gamma_0\right],
\label{Vphi}
\end{equation}
where $V_0$, $\mu$, and $\gamma_0$ are constants, and $p=3,2,4/3,1,2/3$. The modulation contains the energy scale $\Lambda$ and the axion decay constant $f$, both of which are in general functions of $\varphi$. The cosmological phenomena derived from this kind of inflation models, such as the tensor-to-scalar ratio, and the resonantly enhanced modulations of the scalar power spectrum and bispectrum due to the sinusoidal modulations of the potential, have been studied~\cite{cosmology}.

The axion is expected to be coupled to some gauge field via a pseudoscalar-vector coupling,
\begin{equation}
  \mathcal{L}_{\mathrm{int}} = - \frac{\alpha}{4 f}\, \varphi \, \tilde{F}^{\mu\nu} \,F_{\mu\nu},
\label{int}
\end{equation}
where $\alpha$ is a dimensionless coupling constant, $F_{\mu\nu} = \partial_\mu A_\nu - \partial_\nu A_\mu$ is the vector field strength tensor, and $\tilde{F}^{\mu\nu} = \epsilon^{\mu\nu\alpha\beta} F_{\alpha\beta}/2$ is its dual. 
Recently, there has been a lot of studies on the phenomenological effects of this coupling in the axion inflation model. The coupling leads to the particle production with a rate proportional to the rolling speed of axions, which induces a new source of metric perturbations while backacting to the axion dynamics. 
This results in very interesting effects such as the generation of non-Gaussian and non-scale-invariant scalar power spectrum~\cite{barnaby,meer}, 
the production of GWs that could be detected at the ground-based gravity-wave interferometer~\cite{barnaby2,cook},
and the formation of primordial black holes (PBHs) with masses $\lesssim 10^8$g near the end of inflation~\cite{linde,lin,cheng}.
More recently, the authors in Refs.~\cite{cheng2,juan} have made the axion potential steeper locally, or involved non-minimal coupling to gravity~\cite{domcke} to boost the particle production rate at certain wavenumbers in order to seed PBHs with much higher masses. Furthermore, the strength of the GWs associated with the formation of these PBH seeds may reach the sensitivity of future pulsar timing arrays and interferometry experiments.

The cosmic inflation is the most efficient way to seed the formation of PBHs. There have been many inflationary models that produce PBHs with various masses and associated GWs. In a single-field slow-rolling inflation model, matter density perturbations are generally well below the threshold to form PBHs, though they can be formed at rare density peaks. Modifications of the inflation potential to achieve blue-tilted matter power spectra or running spectral indices may lead to large enough density perturbations at the end of inflation~\cite{drees}. However, the resulting PBH masses in most of these models are many orders of magnitude below $M_\odot$~\cite{drees}. To boost the PBH mass into the astrophysical and even the cosmological mass scales, several scenarios involving multi-field inflations have been proposed, such as the hybrid inflation~\cite{hybrid}, the double inflation~\cite{double}, and the curvaton models~\cite{curvaton}, in which small-scale density perturbations can be inflated to a scale ranging from the size of a stellar-mass PBH to that of a supermassive PBH. 

In this paper, we study the formation of PBHs in the axion monodromy inflation governed by the potential~(\ref{Vphi}) and the interaction~(\ref{int}). We will not consider a specific axion model; instead, we treat the model parameters as free as possible. Then, we will probe the parameter space favorable to the formation of PBHs. At the end, we will discuss briefly the physically motivated model with the favorable model parameter. The paper is organized as follows. In the next section, we lay out our base axion inflation model. In Sec.~\ref{scalarpert}, we calculate the curvature perturbation induced by particle productions and then consider the formation of PBHs. In Sec.~\ref{tensorpert}, GWs sourced by the particle production are computed. Section~\ref{conclusion} is our conclusion.

\section{The model}

We consider the axion monodromy inflation in which the axion couples to a $U(1)$ gauge field via the interaction~(\ref{int}).
The action is given by
\begin{equation}
  \mathcal{S} =  \int d^4 x \sqrt{-g} \left[ \frac{M_p^2}{2} \, R  -\frac{1}{2}\partial_\mu\varphi \partial^\mu\varphi- V(\varphi) - \frac{1}{4}F^{\mu\nu}F_{\mu\nu} - \frac{1}{\sqrt{-g}}\frac{\alpha}{4 f} \varphi  \, \tilde{F}^{\mu\nu} \, F_{\mu\nu} \right],
\label{action}
\end{equation}
where $R$ is the curvature scalar and $M_p=2.435\times 10^{18}$ GeV is the reduced Planck mass. For the inflaton potential~(\ref{Vphi}), we choose a linear or concave potential ($p=1$ or $p=2/3$) because convex potentials ($p>1$) are disfavored by cosmic microwave background (CMB) experiments~\cite{planckinflation}, assuming that the ground state is at $\varphi=0$, and taking the approximation that $\Lambda$ and $f$ are constants. Hence, we have
\begin{equation}
V(\varphi)= M_p^p\mu^{4-p} c^{-p}\left[\sqrt{1+\left(\frac{c\varphi}{M_p}\right)^{2p}} - 1\right] +\Lambda^4 \left[1+\cos\left(\frac{\varphi}{f}+\pi\right)\right],
\label{vmodulation}
\end{equation}
where we have introduced a numerical factor $c$ to adjust the smoothness of the potential at $\varphi=0$.
The second term in the above equation is a standard axion potential that has been widely used in axion inflation models as mentioned in the Introduction. The drifts in $\Lambda$ and $f$ during the course of inflation occur in a broad range of axion monodromy scenarios in string theory~\cite{flauger}. We have neglected the drifts and assumed constant $\Lambda$ and $f$. This is a good approximation in the context of the production of PBHs as long as the drifts $\Delta\Lambda,\Delta f \ll M_p$.  It would be interesting to include any possible drifts in the future work.

Here we assume a spatially flat Friedmann-Robertson-Walker metric:
\begin{equation}
ds^2=-g_{\mu\nu} dx^\mu dx^\nu= a^2(\eta) (d\eta^2- d \vec{x}^2),
\end{equation}
where $a(\eta)$ is the cosmic scale factor and $\eta$ is the conformal time related to the cosmic time by $dt=a(\eta)d\eta$.
The Hubble parameter is defined by $H \equiv (da/dt) / a$ or ${\cal H} \equiv (da/d\eta) / a$.

From the action~(\ref{action}), we can write down the Friedmann equation, the equation of motion for the inflaton, and the Maxwell equations, respectively:
\begin{eqnarray}
&& {\cal H}^2 = \frac{1}{3 M_p^2} \left[ \frac{1}{2} \left(\frac{\partial \varphi}{\partial \eta}\right)^2 + \frac{1}{2} \left( \vec{\nabla}  \varphi \right)^2   + a^2 \, V(\varphi) + \frac{a^2}{2} \left( \vec{E}^2 + \vec{B}^2 \right) \right],\label{G00}\\
&& \frac{\partial ^2\varphi}{\partial \eta^2} + 2 {\cal H} \frac{\partial \varphi}{\partial \eta} - \vec{\nabla}^2 \varphi  + a^2 \, \frac{d V}{d \varphi}  = a^2 \frac{\alpha}{f}  \vec{E}\cdot \vec{B},\label{varphieom}\\
&& \frac{\partial^2 \vec{A}}{\partial \eta^2} - \vec{\nabla}^2 \vec{A}
+ \vec{\nabla} (\vec{\nabla}\cdot \vec{A}) = \frac{\alpha}{f}
\frac{\partial \varphi}{\partial \eta} \vec{\nabla} \times \vec{A} - \frac{\alpha}{f} \vec{\nabla} \varphi \times
\frac{\partial \vec{A}}{\partial \eta},\label{maxwell1}\\
&&\frac{\partial}{\partial \eta}(\vec{\nabla} \cdot \vec{A}) = \frac{\alpha}{f}\vec{\nabla} \varphi \cdot(\vec{\nabla} \times \vec{A}),
\label{maxwell2}
\end{eqnarray}
where for the Maxwell equations we have chosen the temporal gauge, i.e. $A_{\mu} = (0, \vec{A})$, and we have introduced the physical ``electric'' and ``magnetic'' fields,
\begin{equation}
\vec{E} = -{1\over a^2} \frac{\partial \vec{A}}{\partial \eta},\quad \vec{B} = {1\over a^2}\vec{\nabla} \times \vec{A}.
\end{equation}

In Ref.~\cite{cheng}, we have calculated the production of gauge quanta by the rolling inflaton via the interaction during a slow-roll inflation, taking into account self-consistently the backreaction of the gauge quanta production on inflation.  To calculate the production of gauge quanta, we separate the inflaton into a mean field and its fluctuations:
\begin{equation}
  \varphi = \phi (\eta) + \delta \varphi (\eta,{\vec x}).
\end{equation}
Under the linear approximation, we decompose ${\vec A}(\eta,{\vec x})$ into two circularly polarized Fourier modes, $A_\pm (\eta,{\vec k})$, which satisfy
the equation of motion as
\begin{equation}
  \left[ \frac{d^2}{d\eta^2} + k^2 \mp 2aH k\xi\right] A_{\pm}(\eta,k) = 0, \;\;\xi \equiv \frac{\alpha}{2 f H}\frac{d\phi}{dt}\,.
\label{photoneom}
\end{equation}
This implies that either one of the two modes, when satisfying the condition $k/(aH)< 2|\xi|$ for the spinoidal instability, grows exponentially fast.
The energy density and the interaction term of the produced gauge quanta are given by the vacuum expectation values of the electric and magnetic fields, respectively,
\begin{eqnarray}
\frac{1}{2}\langle \vec{E}^2+\vec{B}^2 \rangle\hspace{-1mm}&=&\hspace{-1mm}\int\frac{dk\,k^2}{4 \pi^2 a^4} \sum_{\lambda=\pm}\left( \left\vert \frac{dA_\lambda}{d\eta} \right\vert^2 + k^2 \vert A_\lambda \vert^2 \right)\hspace{-1mm}, \\
\langle \vec{E} \cdot \vec{B} \rangle &=& -\int\frac{dk\,k^3}{4 \pi^2 a^4} \frac{d}{d \eta} \left(\vert A_+ \vert^2-\vert A_- \vert^2\right).
\end{eqnarray}
As a consequence, the production of gauge quanta gives rise to a backreaction on the background, whose evolution is then governed by
\begin{eqnarray}
  &&  \frac{d^2\phi}{dt^2} + 3 H \frac{d\phi}{dt} +\frac{dV}{d\phi} = \frac{\alpha}{f} \langle \vec{E}\cdot \vec{B} \rangle, 
\label{inflatoneom} \\
  && 3 H^2 = \frac{1}{M_p^2} \left[ \frac{1}{2}\left(\frac{d\phi}{dt}\right)^2 + V(\phi) + \frac{1}{2} \langle \vec{E}^2 + \vec{B}^2 \rangle \right].
\label{Heom}
\end{eqnarray}

\section{Curvature perturbation and primordial black holes}
\label{scalarpert}

It is well known that the vacuum quantum fluctuation of an inflaton field during the inflation gives rise to the curvature perturbation whose power spectrum is governed by~\cite{olive}
\begin{equation}
\Delta_{\zeta\,{\rm vac}}^2(k)\equiv \langle\zeta(x)^2\rangle=\frac{1}{4\pi^2}\frac{H^4}{(d\phi/dt)^2}.
\label{vacuumDelta}
\end{equation}
In this work, fluctuations of the gauge quanta production during the inflation would lead to a new source for inflaton fluctuations. 
The induced inflaton perturbation satisfies an inhomogeneous equation:
\begin{eqnarray}
\left[ \frac{\partial^2}{\partial t^2} +3 \beta H \frac{\partial}{\partial t} -\frac{{\vec\nabla}^2}{a^2} \hspace{-2mm} \right. &+& \hspace{-2mm} \left. \frac{d^2V}{d\phi^2} \right] \delta\varphi(t,{\vec x}) \nonumber \\ 
&=& {\alpha\over f}\left(  \vec{E}\cdot\vec{B} - \langle \vec{E}\cdot\vec{B} \rangle \right), 
\label{inflatonperteq}
\end{eqnarray}
where the frictional term can be derived as~\cite{anber2,barnaby2}
\begin{equation}
\beta\equiv 1-2\pi\xi \frac{\alpha}{f} \frac{\langle \vec{E}\cdot\vec{B} \rangle}{3H(d\phi/dt)}.
\end{equation}
The particular solution to this equation can be well approximated by~\cite{barnaby2,linde}
\begin{equation}
\delta\varphi= {\alpha\over{3\beta f H^2}}\left(  \vec{E}\cdot\vec{B} - \langle \vec{E}\cdot\vec{B} \rangle\right),
\end{equation}
which contributes to the power spectrum of the curvature perturbation an amount given by
\begin{equation}
\Delta_\zeta^2(k)=\frac{H^2 \langle\delta\varphi^2\rangle}{(d\phi/dt)^2}
=\left[\frac{\alpha\langle \vec{E}\cdot\vec{B} \rangle}{3\beta f H(d\phi/dt)}\right]^2.
\label{Deltazeta}
\end{equation}

PBHs can be arisen from the collapse of over-dense regions originating in the curvature perturbation spawned by the inflaton as they re-entering the horizon in the subsequent expanding universe. If a PBH seed is created during the inflation, the mass of the PBH can be estimated as follows. 
The energy contained within the comoving seed volume that leaves the horizon $N$ e-foldings before the end of inflation is given by
\begin{equation}
{4\over 3}\pi H^{-3} e^{3N} \rho_e~~~{\rm with}~ \quad \rho_e=3H_e^2 M_p^2,
\end{equation}
where $H=H(N)$ is the Hubble parameter at $N$ e-foldings before the end of inflation and $H_e$ at the end of inflation. Let $a_0=1$ be the scale factor at the onset of the inflation, which will eventually span $N_0$ e-foldings over the entire inflationary course. After the inflation has terminated, the universe reheats and is becoming radiation-dominated. Whenever the conformal time $\eta>\eta_e$, the radiation dominant universe would expand to a size of 
\begin{equation}
a(\eta)=H_e e^{2N_0} (\eta-2\eta_e),\quad{\rm where}\quad\eta_e=-(H_e e^{N_0})^{-1}.
\end{equation} 
The comoving volume re-enters the horizon when its scale $k=H e^{N_0-N}$ satisfies the condition $k\eta\sim 1$, i.e., 
when $a=e^{N_0+N} (H_e/H)$ or the temperature of the thermal bath is red-shifted by a factor of $e^N (H_e/H)$~\cite{ng}. 
Therefore, the mass of the PBH presumably formed at this time is 
\begin{equation}
M_{\rm BH}=\frac{4 \pi M_p^2 H_e}{H^2}\,e^{2N}=2.74\times 10^{-38} e^{2N} \left(\frac{M_p H_e}{H^2}\right) M_\odot.
\label{mbh}
\end{equation}

For the purposes of forming PBHs in certain mass ranges associated with GWs of observational interest, we have worked out three specific cases for the potential $V(\phi)$. Figure~1 shows the first case. In this figure and hereafter, we rescale all dynamical variables in terms of the reduced Planck mass. With this inflaton potential, the background solutions for $\phi$ and $\xi$ in Eqs.~(\ref{inflatoneom}) and (\ref{Heom}) are plotted in Fig.~2, for the given initial position and speed of the inflaton, $\phi_0$ and $(d\phi/dt)_0$, and the coupling constant $\alpha$.
The number of e-foldings after the onset of the inflation is defined by $\int_0^t H(t') dt'$. Note that $a_0=1$ and we find that $N_0\simeq 61$. 
Also, in Fig.~3, we have evaluated the scalar spectral index $n_s = 1 - 2\epsilon_1 - \epsilon_2$, the tensor-to-scalar ratio $ r = 16\epsilon_1$, and the running of the scalar spectral index $dn_s/d\ln k$, where the slow-roll parameters $\epsilon_1$ and $\epsilon_2$ are derived from
\begin{equation}
\epsilon_1= - \frac{1}{H^2} \frac{dH}{dt},\quad
\epsilon_2= \frac{1}{H\epsilon_1} \frac{d\epsilon_1}{dt}.
\end{equation}
Figure~4 presents the total power spectrum of the curvature perturbation, which is the sum of the vacuum contribution in Eq.~(\ref{vacuumDelta}) and the induced power in Eq.~(\ref{Deltazeta}). 
Note that the amplitude of the total power spectrum of the curvature perturbation at the largest scales as well as the levels of the scalar spectral index, the tensor-to-scalar ratio, and the scalar spectral index running for cosmologically interesting scales that correspond to about the first $7$ e-foldings are all consistent with the Planck measurements: 
$\Delta_\zeta^2\simeq 2\times 10^{-9}$, $n_s\simeq 0.97$, $r<0.1$, and $|dn_s/d\ln k|<0.013$~\cite{planckparameter}. 

The perturbation $\zeta$ in Eq.~(\ref{Deltazeta}) is obviously non-Gaussian. In Ref.~\cite{linde}, since ${\vec A}$ is a Gaussian vector field, it was argued that 
$\zeta$ can be written as $\zeta=g^2-\langle g^2\rangle$, with $g$ a Gaussian field. Let $P(g)$ be its Gaussian distribution function. Then, the probability distribution function of $\zeta$ is given by $P(\zeta)d\zeta=P(g)dg$, with $g_c^2=\zeta_c+\langle g^2\rangle$, where $\zeta_c\sim 1$ denotes the critical value leading to the PBH formation. This value of $g_c$ for a Gaussian field determines the energy fraction that can collapse to form horizon-sized PBHs with mass $M$ at the re-entry:
\begin{equation}
\beta(M)=\int_{\zeta_c}^\infty P(\zeta)d\zeta = \int_{g_c}^\infty P(g)dg.
\end{equation} 
The astrophysical and cosmological constraints on the PBH energy fraction $\beta(M)$ can then be translated into an upper bound on the power spectrum of the curvature perturbation~\cite{linde}, as shown by the short-dashed line in Fig.~4.

\begin{figure}[htp]
\centering
\includegraphics[width=0.8\textwidth]{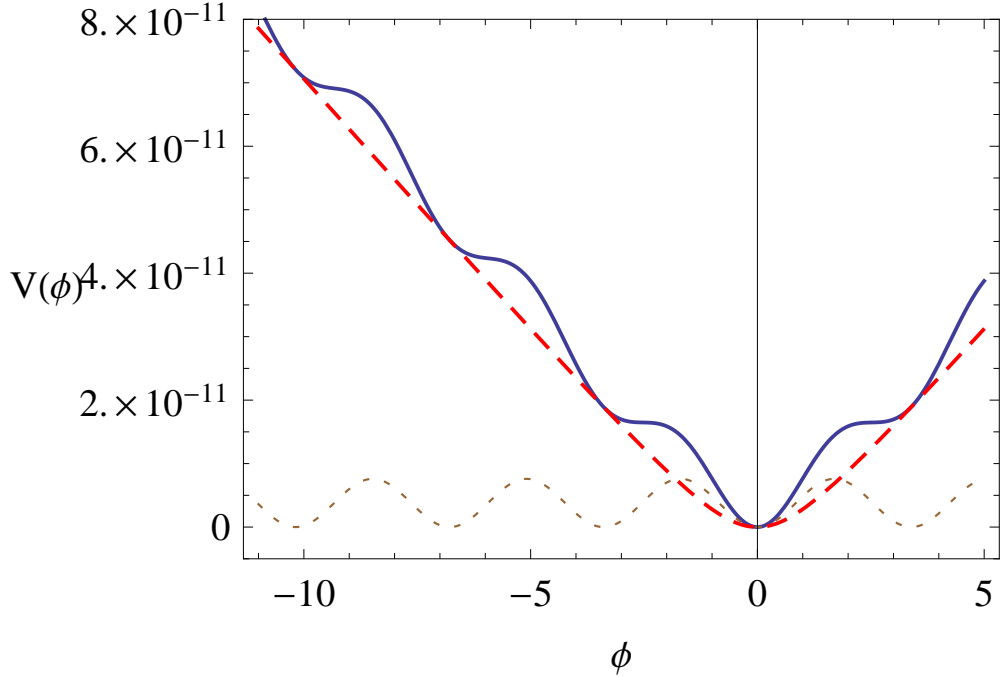}
\caption{All dynamical variables in this figure and in the following figures are rescaled by the reduced Planck mass, $M_p=2.435\times 10^{18}$ GeV. Solid curve denotes the inflaton potential $V(\phi)$ with $p=1$, $c=0.8$, $\mu=2.0\times 10^{-4}$, $f=0.54$, and $\Lambda=1.40\times 10^{-3}$. The dashed and short-dashed curves denote the linear term and the modulation, respectively.}
\label{fig1}
\end{figure}

\begin{figure}[htp]
\centering
\includegraphics[width=0.8\textwidth]{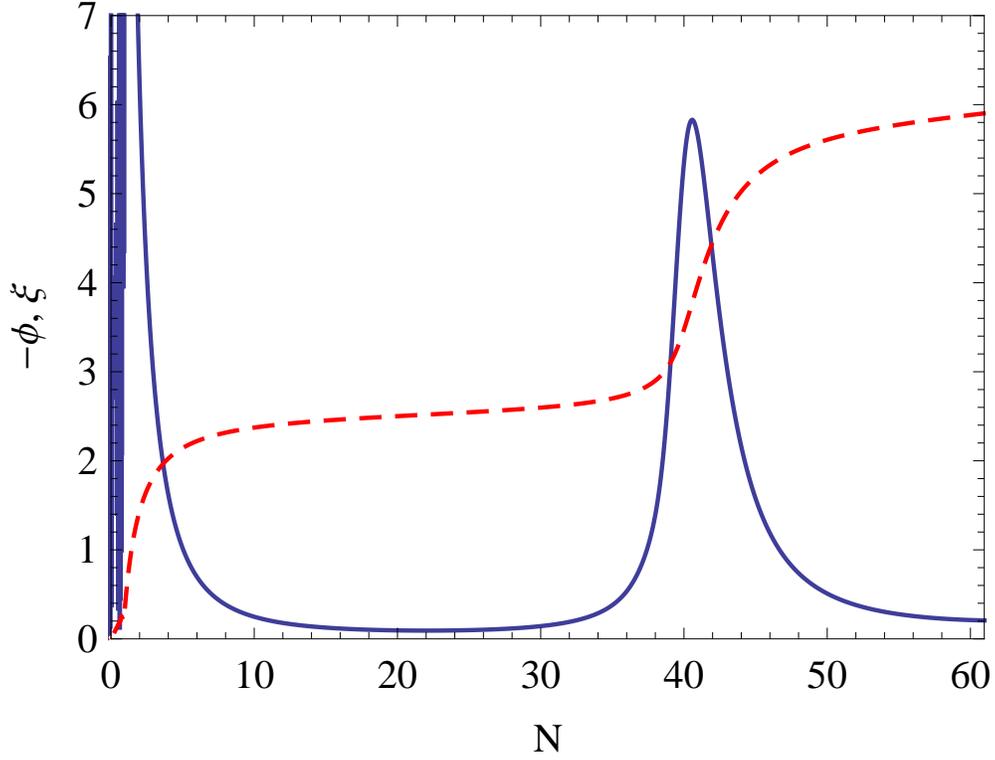}
\caption{Evolutions of $\phi$ (dashed line) and $\xi$ (solid line) against e-foldings $N$ before the end of the inflation for $V(\phi)$ in Fig.~1, with $\phi_0=-5.9$, $(d\phi/dt)_0=7.1 \times 10^{-8}$, and $\alpha=11.8$. The duration of the inflation is $60$ e-foldings.}
\label{fig2}
\end{figure}

\begin{figure}[htp]
\centering
\includegraphics[width=0.8\textwidth]{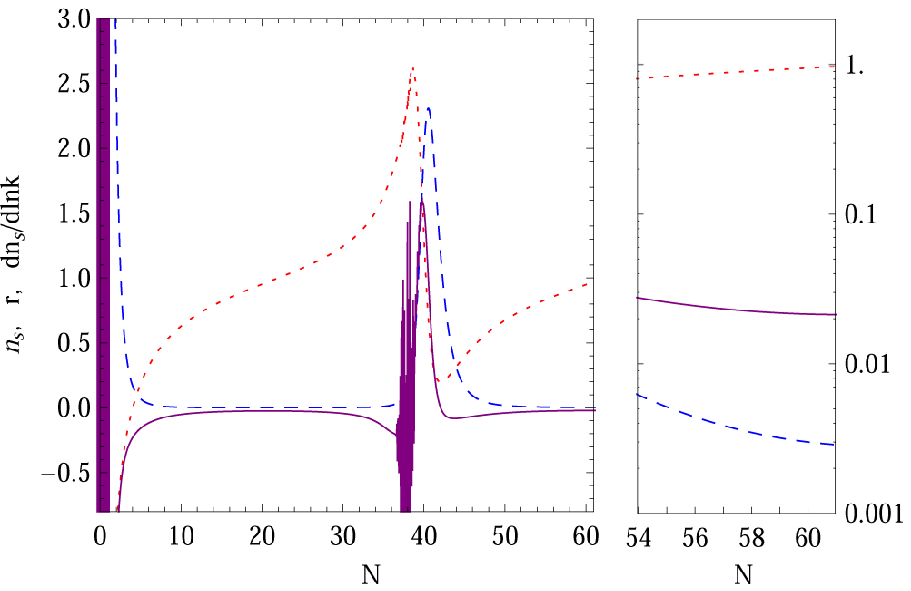}
\caption{Evolutions of $n_s$ (dotted line), $r$ (dashed line), and $dn_s/d\ln k$ (solid line) for $V(\phi)$ in Fig.~1. The right panel zooms in on the first $7$ e-foldings, drawn with $|dn_s/d\ln k|$. }
\label{fig3}
\end{figure}

\begin{figure}[htp]
\centering
\includegraphics[width=0.8\textwidth]{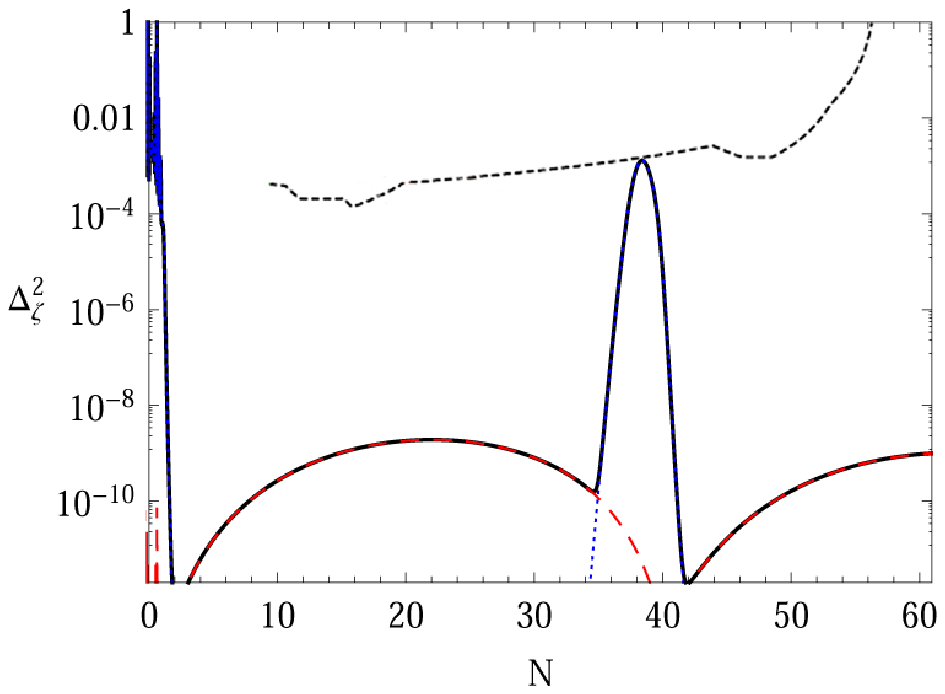}
\caption{The solid line is the total power spectrum of the curvature perturbation for $V(\phi)$ in Fig.~1. The induced and vacuum contributions are denoted by the dotted and dashed lines, respectively. Short-dashed line is the upper bound derived from the astrophysical and cosmological constraints on the PBH abundance.}
\label{fig4}
\end{figure}

In Fig.~2, the first $10$ e-foldings of the inflation are in standard slow-roll stage. When the inflaton enters the steeper slope, it speeds up and hence $\xi$ reaches a maximum value of $5.8$ at $N\simeq 40$. This produces a huge amount of gauge quanta and induces a high peak of inflaton fluctuations at $N\simeq 38.5$. Consequently, the strong backreaction due to the gauge field production almost stops the inflaton motion at $\phi\simeq -2.6$ for about half of the inflationary duration. After then, the inflaton speeds up again and the inflation ends up with a rather complicated dissipation-fluctuation processes. This is a typical example of the so-called trapped inflation in which the production of gauge quanta can sustain a nearly steady thermal bath during the inflationary epoch and exhaust the vacuum energy to terminate the inflation gracefully without undergoing a large-scale preheating or perturbative reheating~\cite{cheng}.

In Ref.~\cite{perturbativity}, the authors have shown that in the in-in formalism the validity of the perturbative treatment of axion and gauge modes restricts $\xi\lesssim 5$. However, this result does not apply to our present work. Instead of using the perturbation approach, we have taken only the linear approximation for the axion and gauge modes in the equations of motion to carry out all calculations including the strong backreaction. The validity of this linear approximation was recently discussed in Ref.~\cite{fugita} for the same gauge-field axion inflation, in which the authors have proposed a consistency condition for the linearity. This condition was later analyzed in a detailed numerical calculation by including the backreaction self-consistently, which shows that the linearity condition can be maintained for strong couplings as large as $\xi\lesssim 10$~\cite{cheng}. In Fig.~2, our main concern is the peak at $N\simeq 40$, where $\xi\simeq 5.8$ is well within the linearity condition. Near the end of the inflation $\xi$ gets larger values that drive the scalar power spectrum to high spikes at which PBHs of much smaller masses are likely to be copiously formed. However, these high-density spikes should be damped to a lower level because of the local strong gravity. This gravity effect has been ignored in the present work; indeed, it can be properly taken into account by including a higher-order gravitational term in the inflaton perturbation Eq.~(\ref{inflatonperteq})~\cite{fugita}. From Eq.~(\ref{mbh}), a black hole with $M_{\rm BH}\simeq 19.7 M_\odot$ can be produced by taking $H_e=1.48\times 10^{-6}$ and $H=2.38\times 10^{-6}$ when $N=38.5$. Hence, the peak of the scalar power spectrum in Fig.~4 will seed the formation of higher solar-mass PBHs,
with the black hole fraction $\beta(M)\sim 10^{-9}$, which gives the fraction of dark matter in PBHs $f(M)\sim 0.1$~\cite{linde,carr16}. In Appendix, we will present two other inflaton potentials that can produce PBHs with smaller masses.

\section{Associated gravitational waves}
\label{tensorpert}

In addition to de Sitter vacuum fluctuations, GWs can be directly sourced by the gauge field production~\cite{barnaby,sorboGW,cheng2,juan}. The GW equation reads 
\begin{equation}
\left[\frac{\partial^2}{\partial \eta^2} + {2\over a}\frac{da}{d\eta} \frac{\partial}{\partial \eta} -{\vec\nabla}^2\right] h_{ij}
= \frac{2a^2}{M_p^2}\left( -E_i E_j - B_i B_j \right)^{TT}, 
\label{sourcedGW}
\end{equation}
where $TT$ denotes the transverse and traceless projection of spatial components of the energy-momentum tensor of the gauge field. 

While the curvature perturbation is sufficiently large for seeding the formation of PBHs, GWs simultaneously produced at the horizon crossing as the second-order effect in the metric perturbation theory cannot be neglected~\cite{inomata,juan}. The second-order effect is mainly contributed by the transverse-traceless part of a source term $S_{ij}(\delta\varphi)$ involving quadratic terms of the curvature perturbation that should appear on the right-hand side of the GW Eq.~(\ref{sourcedGW}). In the present consideration, we will show that the amount of GWs sourced by $S_{ij}(\delta\varphi)$ is subdominant to that by the gauge field production. From naive power counting, the GW amplitude induced by the gauge field can be estimated as $h_A\sim A^2\sim\zeta$, whereas that induced by $S_{ij}(\delta\varphi)$ is $h_{\delta\varphi}\sim\delta\varphi^2\sim \zeta^2$. As long as $\Delta_\zeta^2\sim\zeta^2$ is beneath the PBH bound, $\zeta\ll 1$ and thus we have $h_{\delta\varphi}\ll h_A$. In fact, this simple estimation is supported by detailed calculations~\cite{cheng,juan}. In Ref.~\cite{cheng}, it was shown that the energy densities of inflaton perturbation $\rho_{\delta\varphi}$ and of gauge quanta $\rho_A$ during inflation scale as $\rho_{\delta\varphi}/\rho_A \sim \zeta \ll 1$. This implies that the energy-momentum tensor that sources the generation of GWs mainly comes from the gauge quanta rather than the inflaton perturbation. More recently, the authors in Ref.~\cite{juan} have confirmed the sub-dominance of the second-order effect by explicitly computing the GW power spectra induced by the gauge field production as well as by the second-order curvature perturbation.

Rather than numerically solving Eq.~(\ref{sourcedGW}) in conjunction with the gauge mode equation~(\ref{photoneom}), we use the approximate analytic gauge mode solutions to estimate the present relative GW energy density per logarithmic $k$ interval, given by~\cite{barnaby,sorboGW}
\begin{equation}
\Omega_{\rm GW} h^2\simeq 3.5\times 10^{-7} \frac{H^2}{M_p^2} \left(1+4.3\times 10^{-7}\frac{H^2}{M_p^2} \frac{e^{4\pi\xi}}{\xi^6}\right).
\end{equation}
To evaluate $\Omega_{\rm GW} h^2$, we treat $\xi$ as a function of $N$ given by the numerical results plotted as $\xi(N)$ in Figs.~2, 7, and 11. The Hubble scale $H=H(N)$ is calculated by Eq.~(\ref{Heom}). This spectral energy density is at $k=H e^{N_0-N}$ with $H_0$ corresponding to the present horizon of size $0.002\, {\rm Mpc}^{-1}$.  In Fig.~5, we plot $\Omega_{\rm GW} h^2$ against the frequency $f=k/(2\pi)=3\times 10^{-18} (H/H_0) e^{N_0-N} {\rm Hz}$, where $N_0$ is the total number of e-foldings for the inflationary duration with inflaton potentials plotted in Figs.~1, 6, and 10, respectively. This gives rise to three prominent peaks, which from left to right are associated with the production of PBHs with masses of $19.7 M_\odot$, $2.4\times 10^{-13} M_\odot$, and  $1.2\times 10^8 {\rm g}$, respectively. In the figure, we list the current upper limits on GW background inferred from pulsar timing array data~\cite{PTA} and aLIGO O1 data~\cite{ligoback17}. Also shown are the projected sensitivities of on-going and future GW experiments such as aLIGO O3, O5~\cite{ligoback16}, LISA~\cite{lisa}, and SKA radio telescope~\cite{ska}.

\begin{figure}[htp]
\centering
\includegraphics[width=0.8\textwidth]{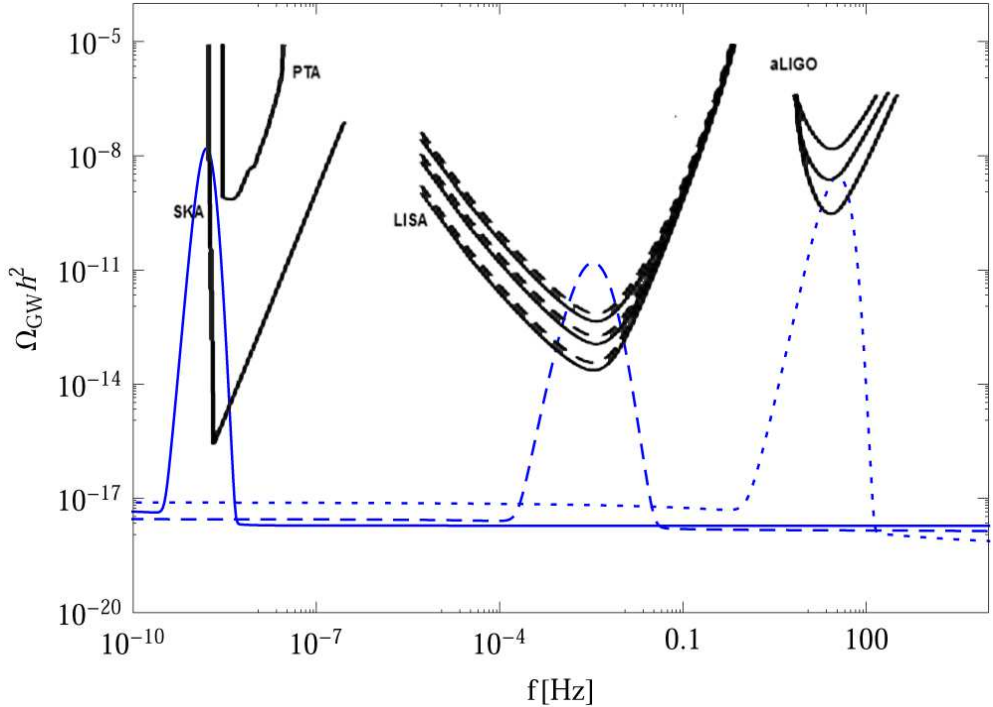}
\caption{Peaks from left to right represent the present spectral energy density of the gravitational waves associated with the production of PBHs with masses of $19.7 M_\odot$, $2.4\times 10^{-13} M_\odot$, and  $1.2\times 10^8 {\rm g}$, respectively. Also shown are the upper limits set by the pulsar timing array experiments (PTA) and aLIGO O(1) (the uppermost curve), as well as the projected aLIGO O3, O5, LISA, and SKA sensitivities.}
\label{fig5}
\end{figure}

\section{Conclusion}
\label{conclusion}

The axion monodromy inflation is a well-motivated inflationary scenario in string theory. The inflation is driven by the axion with super-Planckian field values in a monomial potential with superimposed sinusoidal modulations. While the monomial potential provides a slow-roll inflation consistent with CMB data and large-scale-structure surveys, the modulations may give rise to interesting observational signatures. 

Flauger {\it et al.} in Ref.~\cite{cosmology} have found that the periodic modulations can drive resonant enhancements of inflaton perturbations, with characteristic scale-dependent modulated amplitude, giving rise to oscillations in the CMB anisotropy power spectrum. For an axion decay constant $f\ll M_p$, they have calculated the modulated curvature power spectrum, and determined the limits that CMB data places on the amplitude and frequency of modulations. In some favorable axion monodromy models, resonant contributions to the CMB power spectrum and bispectrum could be detected in near-future CMB experiments.

In this paper, we have studied the consequences from the periodic modulations of the axion potential with $f\sim M_p$, when the axion is coupled to a massless gauge field with a coupling constant $\alpha\sim 20$. The modulations with Planckian frequencies have little effect on the slow-roll inflation at CMB scales. After the inflaton leaves the slow-roll regime and slides down a steeper slope of the modulated potential, a huge amount of gauge quanta is produced. As a consequence, the backreaction of the particle production to the inflaton motion induces large inflaton fluctuations, creating a peak in the curvature power spectrum that seeds the formation of primordial black holes. We have given permissible model parameters that produce PBHs with masses ranging from $10^8$ grams to $20M_\odot$. Interestingly, the amplitudes and frequencies of gravitational waves sourced by the stress-energy tensor of the generated gauge quanta could lie within the sensitivities of on-going and future gravity-wave detectors such as aLIGO/VIRGO, LISA, and pulsar timing arrays.

The axion monodromy provides a sound theoretical framework for a large-field inflation model to work with. In addition to the monomial potential that realizes the slow-roll regime, inherent nonperturbative effects generate small sinusoidal modulations of the potential with model-dependent axion decay constant $f$ and energy scale $\Lambda$. In the present consideration, we have introduced an axion-photon interaction with a coupling constant $\alpha$. This interaction strength can be rewritten as $\alpha/f=\alpha_e/(2\pi f_a)$, where $\alpha_e\simeq 1/137$ is the fine structure constant and the rescaled axion decay constant $f_a\sim 10^{-4} f$. It is expected that the axion decay constant $f$ is in the order of the grand unification scale, which is generally smaller than the Planck scale. However, there are some examples in string theory that allow $f$ to be near the Planck scale~\cite{banks}. The preferred values for the model parameters that can produce primordial black holes and associated gravitational waves of astrophysical interest are $f\sim M_p$, $\Lambda\sim 10^{-3} M_p$, and $f_a\sim 10^{-4} M_p$. These disparate energy scales may be constructed, for example, in the clockwork mechanism~\cite{clockwork}. Overall, the axion monodromy inflation has very rich astrophysical and cosmological implications that may be tested in on-going and future CMB and gravity-wave experiments.

\begin{acknowledgments}
The authors would like to thank K. Choi, M. Peloso, and E. Silverstein for useful conservations.
This work was supported in part by the Ministry of Science and Technology, Taiwan, ROC under Grants No. MOST104-2112-M-001-039-MY3 (K.W.N.) and No. MOST104-2112-M-003-013 (W.L.).
\end{acknowledgments}

\appendix*
\section{Two more cases with smaller PBH masses}

Figure~6 and Fig.~10 are the second and third potentials, respectively.  Similar to the first case, we have calculated the background and the perturbation in these two potentials. The results are summarized in Figs.~7-9 and Figs.~11-13. Figure~8 and Fig.~12 show $n_s$, $r$, and $dn_s/d\ln k$, all of which satisfy the Planck constraints on CMB scales. In Fig.~7, the $\xi$ bump at $N\simeq 25$ induces a $\Delta_\zeta$ peak at $N\simeq 23$ with a peak value saturating the PBH bound as seen in Fig.~9, which corresponds to $M_{\rm BH}\simeq 2.4\times 10^{-13} M_\odot$, by taking
$H_e=3.99\times 10^{-7}$ and $H=2.15\times 10^{-6}$. 
This contributes to the fraction of dark matter in PBHs of amount $f(M)\sim 0.06$~\cite{linde,carr16}.
In Fig.~11, the $\xi$ peak at $N\simeq 9$ produces a $\Delta_\zeta$ peak just outside the PBH bound as shown in Fig.~13. This density peak at $N\simeq 8$ seeds PBHs with $M_{\rm BH}\simeq 1.2\times 10^8 {\rm g}$ when $H_e=8.54\times 10^{-7}$ and $H=1.78\times 10^{-6}$. These small PBHs would have evaporated through the
emission of Hawking radiation.

\begin{figure}[htp]
\centering
\includegraphics[width=0.8\textwidth]{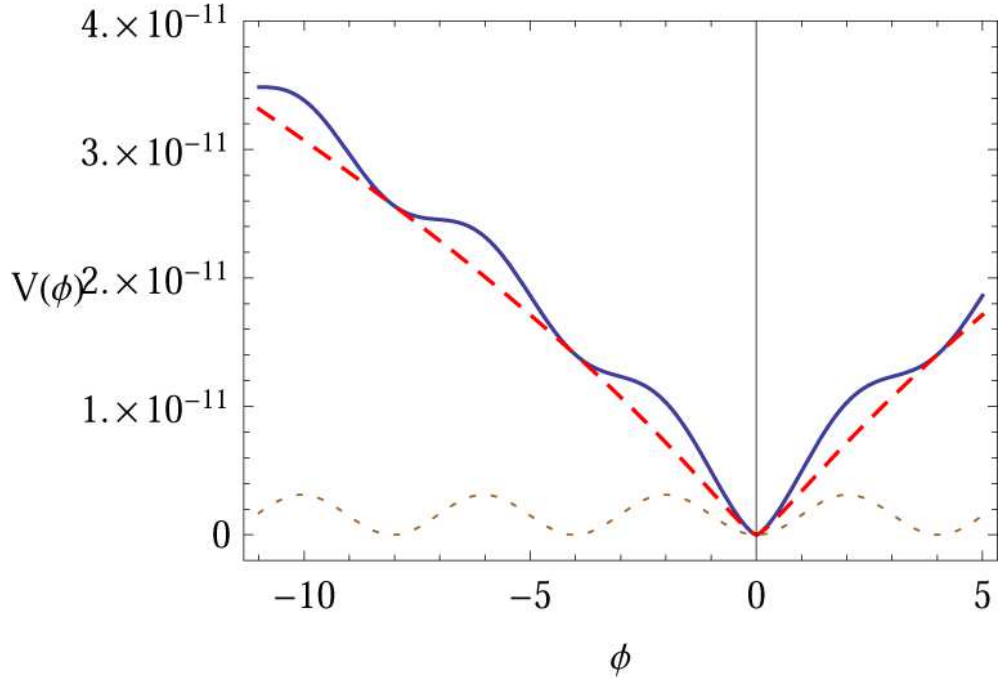}
\caption{The solid curve denotes the inflaton potential $V(\phi)$ with $p=2/3$, $c=1$, $\mu=4.7\times 10^{-4}$, $f=0.64$, and $\Lambda=1.12\times 10^{-3}$. The dashed and short-dashed curves denote the concave term and the modulation, respectively.}
\label{fig6}
\end{figure}

\begin{figure}[htp]
\centering
\includegraphics[width=0.8\textwidth]{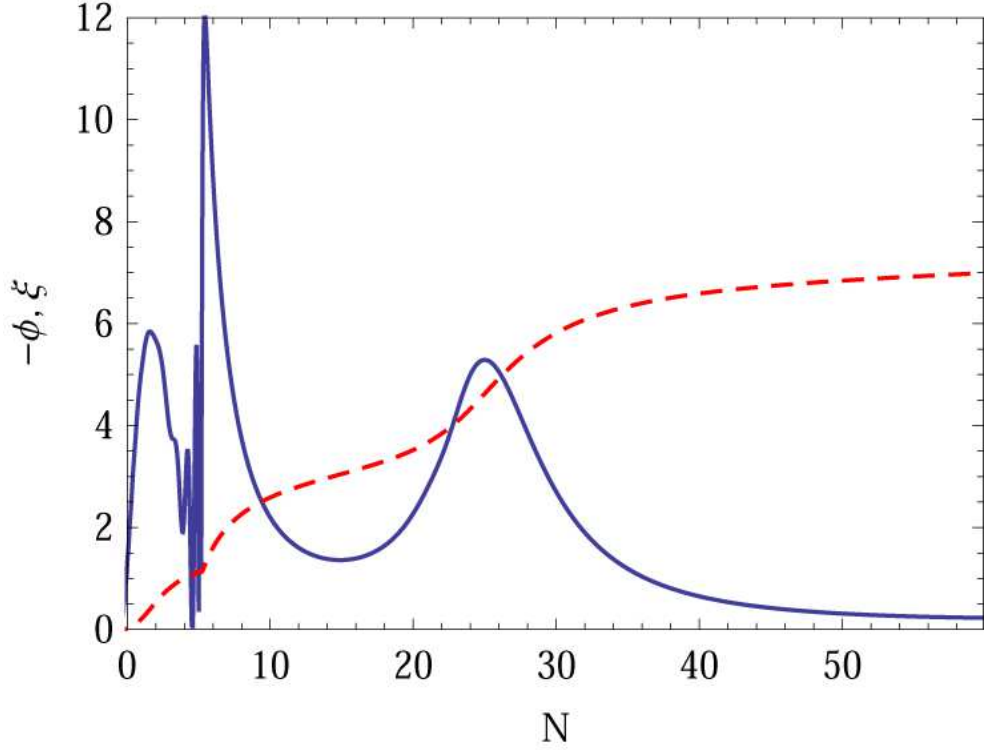}
\caption{Evolutions of $\phi$ (dashed line) and $\xi$ (solid line) for $V(\phi)$ in Fig.~5 with $\phi_0=-7.0$, $(d\phi/dt)_0=3.8\times 10^{-8}$, and $\alpha=22.3$. The inflation lasts for $60$ e-foldings.}
\label{fig7}
\end{figure}

\begin{figure}[htp]
\centering
\includegraphics[width=0.8\textwidth]{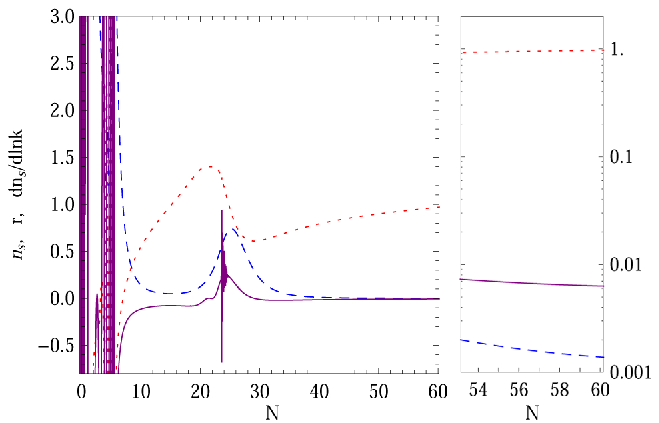}
\caption{Evolutions of $n_s$ (dotted line), $r$ (dashed line), and $dn_s/d\ln k$ (solid line) for $V(\phi)$ in Fig.~5. The right panel zooms in on the first $7$ e-foldings, drawn with $|dn_s/d\ln k|$.}
\label{fig8}
\end{figure}

\begin{figure}[htp]
\centering
\includegraphics[width=0.8\textwidth]{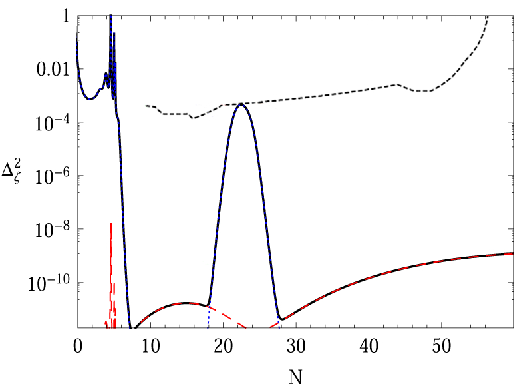}
\caption{The solid line is the total power spectrum of the curvature perturbation  for $V(\phi)$ in Fig.~5. The induced and vacuum contributions are denoted by the dotted and dashed lines, respectively. Short-dashed line is the primordial black hole bound. }
\label{fig9}
\end{figure}

\begin{figure}[htp]
\centering
\includegraphics[width=0.8\textwidth]{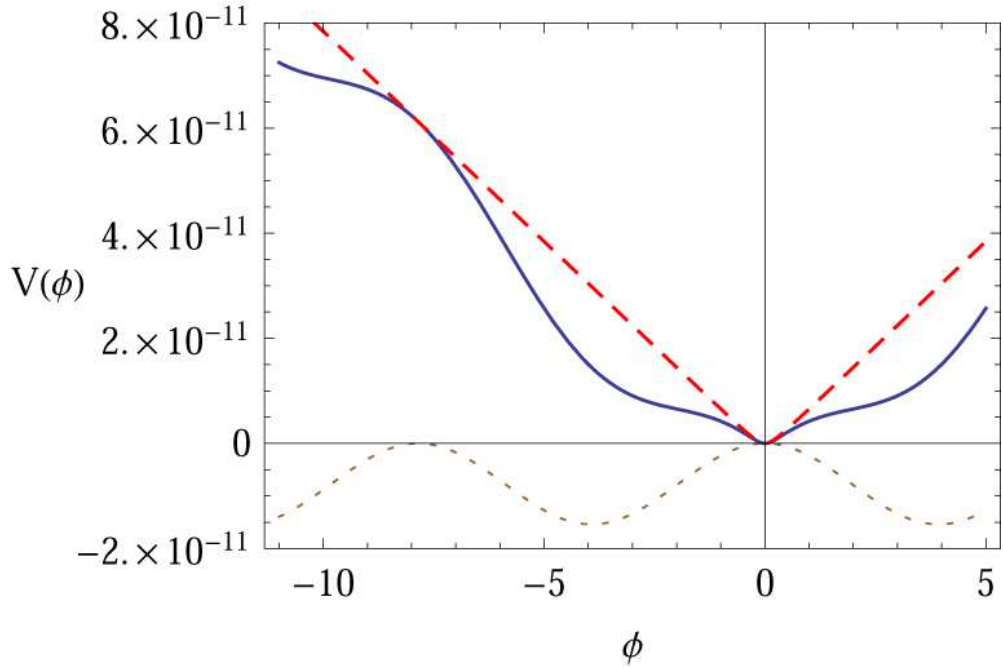}
\caption{The solid curve denotes the inflaton potential $V(\phi)$ with $p=1$, $c=5$, $\mu=2.0\times 10^{-4}$, $f=1.25$, and $\Lambda=1.67\times 10^{-3}$. The dashed and short-dashed curves denote respectively the linear term and the modulation, with a sign change of the modulation term in Eq.~(\ref{vmodulation}).}
\label{fig10}
\end{figure}

\begin{figure}[htp]
\centering
\includegraphics[width=0.8\textwidth]{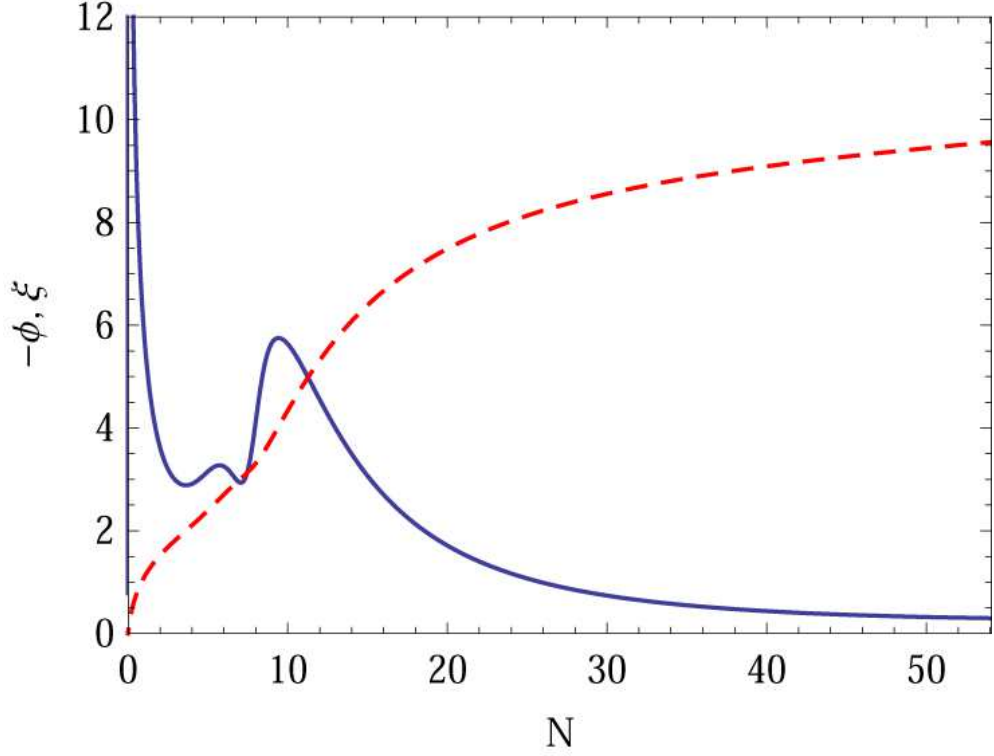}
\caption{Evolutions of $\phi$ (dashed line) and $\xi$ (solid line) for $V(\phi)$ in Fig.~9 with $\phi_0=-9.6$, $(d\phi/dt)_0=1.4\times 10^{-7}$, and $\alpha=26.3$. The inflation lasts for $54$ e-foldings.}
\label{fig11}
\end{figure}

\begin{figure}[htp]
\centering
\includegraphics[width=0.8\textwidth]{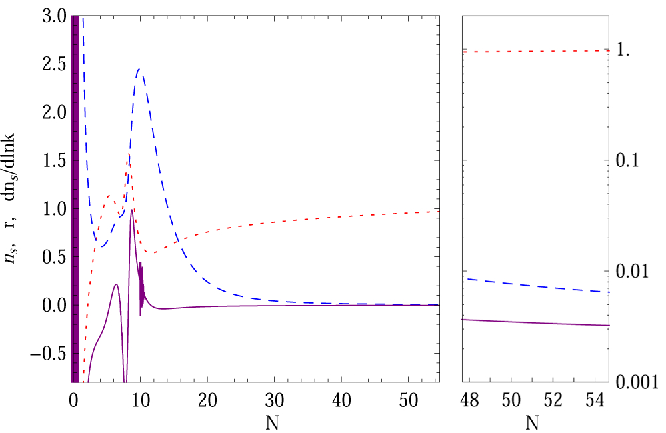}
\caption{Evolutions of $n_s$ (dotted line), $r$ (dashed line), and $dn_s/d\ln k$ (solid line) for $V(\phi)$ in Fig.~9. The right panel zooms in on the first $7$ e-foldings, drawn with $|dn_s/d\ln k|$.}
\label{fig12}
\end{figure}

\begin{figure}[htp]
\centering
\includegraphics[width=0.8\textwidth]{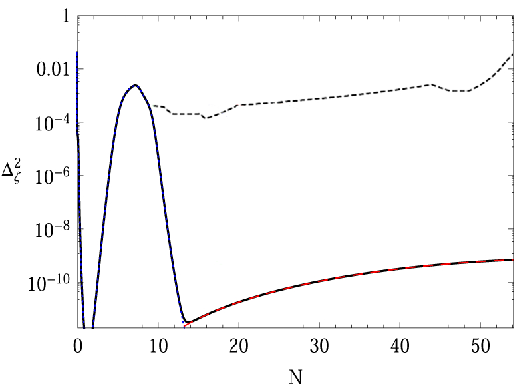}
\caption{The solid line is the total power spectrum of the curvature perturbation  for $V(\phi)$ in Fig.~9. The induced and vacuum contributions are denoted by the dotted and dashed lines, respectively. Short-dashed line is the primordial black hole bound. }
\label{fig13}
\end{figure}

\end{document}